\documentclass[a4paper,11pt,twoside,fleqn]{article}

\usepackage[latin1]{inputenc}
\usepackage[T1]{fontenc}
\usepackage{amsfonts}
\usepackage{amssymb}
\usepackage{latexsym}
\usepackage{longtable}
\usepackage{array}
\usepackage{amsmath}
\usepackage{epsfig}
\usepackage{float}
\usepackage{xy}

\newlength{\amplitude}
\numberwithin{equation}{section}

\makeatletter
\renewcommand{\section}{\@startsection{section}{1}{0mm}   {\baselineskip}{1\baselineskip}{\normalfont\Large\scshape\centering\textbf}}
\makeatother
\makeatletter
\renewcommand{\subsection}{\@startsection{subsection}{2}{0mm}   {\baselineskip}{1\baselineskip}{\normalfont\large\itshape\centering\textbf}}
\makeatother
\makeatletter
\renewcommand{\subsubsection}{\@startsection{subsubsection}{2}{0mm}   {\baselineskip}{1\baselineskip}{\normalfont\normalsize\itshape\centering\textbf}}
\makeatother

\floatstyle{ruled}
\restylefloat{figure} 

\title{\textsc{\textbf{New $\mathcal{SU}(1,1)$ Position-Dependent Effective Mass Coherent States for the Generalized Shifted Harmonic Oscillator}}}

\author {Sid-Ahmed YAHIAOUI\footnote{Email address:          \texttt{s$\_$yahiaoui@univ-blida.dz}}$\quad$and\quad Mustapha BENTAIBA\footnote{Email address: \texttt{bentaiba@univ-blida.dz}}\\[2mm]
        \normalsize\textit{LPTHIRM, Département de Physique, Faculté des Sciences,}\\[-3mm]
        \normalsize\textit{Université Saâd DAHLAB de Blida, B.P. 270 Route de Soumâa, 09000 Blida, Algeria}}

\begin{document}
\maketitle
\noindent\hrulefill\vspace{-3mm}
\begin{abstract}
\noindent A new $\mathcal{SU}(1,1)$ position-dependent effective mass coherent states (PDEM CS) related to the shifted harmonic oscillator (SHO) are deduced. This is accomplished by applying a similarity transformation to the generally deformed oscillator algebra (GDOA) generators for PDEM system and construct a new set of operators which close the $\mathfrak{su}(1,1)$ Lie algebra, being the PDEM CS of the basis for its unitary irreducible representation. The residual potential is associated to the SHO. From the Lie algebra generators, we evaluate the uncertainty relationship for a position and momentum-like operators in the PDEM CS and show that it is minimized in the sense of Barut-Girardello CS. We prove that the deduced PDEM CS preserve the same analytical form than those of Glauber states. We show that the probability density of dynamical evolution in the PDEM CS oscillates back and forth as time goes by and behaves as classical wave packet.
\vspace{1mm}
{\flushleft{\textbf{\small PACS numbers:} 03.65.Fd; 11.30.Pb; 42.50.Dv}\vspace{-2mm}
}\vspace{-6mm}
\end{abstract}
\hrulefill
\maketitle

\vspace{4mm}

\section{Introduction}%

\noindent The use of a varying mass in studying and solving the Schr\"odinger equation has long been left. Since the properties of a quantum system has a close relationship with its dimensionality of space, the study of such system with position-dependent effective mass (PDEM) has been the subject of much activity and many approaches and techniques have been devoted to constructing exactly solvable potentials for PDEM Schr\"odinger equation over the last years~\cite{1,2,3,4,5,6,7,8,9,10,11,12,13,14,15}. They have found important applications in the fields of material science and condensed matter physics such as semiconductors~\cite{16}, $^3$He clusters~\cite{17}, quantum wells, wires and dots~\cite{18}, and quantum liquids~\cite{19}, etc.\\
\indent However, one of the well-known problem of the PDEM consists to define the kinetic energy operator when the mass, say $M(x)$, is a function of position since the momentum and mass operators no longer commute. In order to deal with this difficulty, the ordering ambiguity of the mass and the momentum was addressed by von Roos~\cite{20}
\begin{equation}\label{1.1}
\hat h=\frac{1}{4}\Big(M^a(x)\,\hat p\,M^b(x)\,\hat p\,M^c(x)+M^c(x)\,\hat p\,M^b(x)\,\hat p\,M^a(x)\Big)+V(x),
\end{equation}
which has the advantage of a built in Hermiticity. Here, $\hat p\:\big(\!\equiv-i\hbar\frac{d}{dx}\big)$ is the conventional momentum operator, $M(x)=m_0 m(x)$ is the position-dependent mass function ($m(x)$ being dimensionless) and the parameters $a,\,b$ and $c$ are constrained by the condition $a+b+c=-1$.\\
\indent On the other hand, coherent states (CS), originally introduced by Glauber~\cite{21} in the context of quantum optics to characterize those states of the electromagnetic field that factorize the field coherence function, have been played a crucial role in mathematical physics in the last decades. There are three definitions of CS, and as results, they all lead to the equivalent state vector for the harmonic oscillator: (i) they are eigenstates of the annihilation operator (Glauber's approach) and later generalized by Gazeau and Klauder~\cite{22}, (ii) they are displayed version of the ground wave-function (Klauder's approach) and (iii) they minimize the Heisenberg position-momentum uncertainty relation (Schr\"odinger's approach).\\
\indent The mentioned definitions overlap only for the special case of Heisenberg-Weyl group $H_4$. However, this latter is not the only group for which we can construct coherent states; the notion of generalized CS has been extended by Barut and Girardello~\cite{23}, and Perelomov~\cite{24} in the context of an unitary irreducible representation of any Lie group $\mathcal G$ on the Hilbert space $\mathfrak H$. The basic idea of generalization takes its originality on the fact that the problem becomes more complicated if any quantum system is not connected to a symmetry group.\\
\indent In particular, the most generalized CS which have been investigated in the literature are those associated with groups $\mathcal{SU}(2)$ and $\mathcal{SU}(1,1)$~\cite{25,26,27,28,29,30,31,32,33,34,35,36,37,38}. For the usual $\mathcal{SU}(1,1)$ group, which is the most elementary noncompact and non-Abelian simple Lie group, they are two set of CS but not equivalent, namely: the so-called Barut-Girardello CS which are characterized by the complex eigenvalues $\alpha$ of the noncompact generator $\mathcal{\widehat K_-}$ of the $\mathfrak{su}(1,1)$ algebra
\begin{eqnarray}\label{1.2}
\mathcal{\widehat K_-}|\alpha;k\rangle_{\rm BG}=\alpha|\alpha;k\rangle_{\rm BG},\qquad \textrm{and}\qquad\mathcal{\widehat K_-}|0,k\rangle=0,
\end{eqnarray}
where $|n=0,k\rangle$ is the lowest normalized state and $k$ is the Bargmann index of the representation $\mathcal{SU}(1,1)$, and the Perelomov CS which are characterized by point $\zeta=(\rho/|\rho|)\tanh|\rho|$, so $|\zeta|<1$ and obtained by applying the unitary operator $\Omega(\rho)$ of the coset space $\mathcal{SU}(1,1)/\mathcal U(1)$ to the lowest state $|0,k\rangle$
\begin{eqnarray}\label{1.3}
|\zeta;k\rangle_{\rm P}&=&\exp(\rho\,\mathcal{\widehat K_+}-\rho^\ast \mathcal{\widehat K_-})|0,k\rangle\nonumber\\
&=& (1-|\zeta|^2)^k\,e^{\zeta \mathcal{\widehat K_+}}|0,k\rangle.
\end{eqnarray}
\indent The condition $|\zeta|<1$ shows clearly that the $\mathcal{SU}(1,1)$ Perelomov CS are defined in the interior of the unit disc.\\ \newpage
\indent In the previous paper~\cite{39}, we have succeed to construct a set of PDEM CS for a wide kind of pseudo-Hermitian Hamiltonians as well as for their Hermitian counterpart Hamiltonians under the generally deformed oscillator algebra (GDOA). The main aim of the present paper is to develop the procedure which ensure that all ladder operators associated with a GDOA, say $\big\{\mathcal{\widehat H},\,\mathcal{\widehat Q_\pm}\big\}$, in the Hilbert space $\mathfrak H$ can indeed be transformed, through a similarity transformation, to the generators of the $\mathfrak{su}(1,1)$ Lie algebra. Results above are very interesting since they show that $\mathcal{\widehat Q_\pm}$ and $\mathcal{\widehat H}$ share the same algebraic structure that operators $\mathcal{\widehat K_\pm}$ and $\mathcal{\widehat K}_0$, respectively, to get the unitary irreducible representation for the $\mathfrak{su}(1,1)$ Lie algebra. These generators are connected in the natural way to the PDEM of the shifted harmonic oscillator (SHO). As a result, a new $\mathcal{SU}(1,1)$ PDEM CS for the SHO, in the sense of Barut-Girardello, follow from this construction. We evaluate the position-momentum uncertainty relation in the PDEM CS basis and we show that they preserve the same analytical form than those of Glauber states.\\
\indent The organization of the paper is as follows. In the section 2 we introduce a detailed recipe for obtaining the GDOA from the general procedure for solving PDEM and its structure function is deduced. In section 3 we exploit the analogy with the so-called amplitude-squared realization in quantum optics and we apply the similarity transformation to the GDOA in order to construct three operators which seem to close the $\mathfrak{su}(1,1)$ Lie algebra. In the section 4, the related PDEM CS are constructed as eigenfunctions of the annihilation operator of the $\mathfrak{su}(1,1)$ Lie algebra and we prove that they are of minimim uncertainty, according to the uncertainty relation and are more stable with time evolution. Finally, some concluding remarks are given in the last section.

\section{Position-dependent effective mass and factorization}%

\noindent Let us consider the one dimensional PDEM Schr\"{o}dinger equation. Following Bagchi and coworkers~\cite{1}, setting $M(x)=1/U^2(x)$, where $U(x)$ is some positive-definite function playing the role of deforming function, into Eq.~\eqref{1.1}, this will result in the following time independent wave equation
\begin{eqnarray}\label{2.1}
-\frac{\hbar^2}{4m_0}\bigg(\!U^A(x)\frac{d}{dx}U^B(x)\frac{d}{dx}U^C(x)
+U^C(x)\frac{d}{dx}U^B(x)\frac{d}{dx}U^A(x)+V(x)\!\bigg)\psi(x)=\mathcal{E}\psi(x),
\end{eqnarray}
where the parameters $A,\,B$ and $C$ are constrained by a new condition $A+B+C=2$.\\
\indent Then Eq.~\eqref{2.1} can be reduced to the following form
\begin{eqnarray}\label{2.2}
\hat h\,\psi(x)=\bigg[-\frac{\hbar^2}{2m_0}
\bigg(\sqrt{U(x)}\frac{d}{dx}\sqrt{U(x)}\bigg)^2+V_{\rm eff.}(x)\bigg]\psi(x)
=\mathcal{E}\psi(x),
\end{eqnarray}
where the conventional momentum $\hat p=-i\hbar\frac{d}{dx}$ is substituted by the deformed momentum operator $\hat p\ \rightarrow\ \hat\pi=\sqrt{U(x)}\,\hat p\,\sqrt{U(x)}$.\\
\indent Here the function
\begin{eqnarray}\label{2.3}
  V_{\rm eff.}(x)=V(x)+\frac{\hbar^2}{2m_0}\Big(\frac{1}{2}-A\Big)\Big(\frac{1}{2}-C\Big)U'^2(x)
  +\frac{\hbar^2}{4m_0}\Big(1-A-C\Big)U''(x)U(x),
\end{eqnarray}
plays the role of an effective potential and depends on the ambiguity parameters, the explicit expressions of the dimensionless mass $m(x)$ and the initial potential $V(x)$.\\
\indent Let us look for solutions of Eq.~\eqref{2.2} in the form $\psi(x)=e^{\theta(x)} \phi(x)$~\cite{34} and define a transformation $x\mapsto\xi$ for a mapping $\xi\equiv\mu(x)$. One can now easily reduce Eq.~\eqref{2.2} to the Schr\"{o}dinger equation with constant mass by eliminating the linear differential term, which requires that $\mu(x)=\int^x \mathcal S(y)\,dy$ and $\theta(x)=\ln\mathcal S^{1/2}(x)$, with $\mathcal S(x)=\frac{1}{U(x)}$. Thus, the wave-function read explicitly as\footnote{\;The notation of the Ref.~\cite{34} is kept.}
\begin{eqnarray}\label{2.4}
\psi(x)=\mathcal S^{1/2}(x)\,\phi(x)\equiv\mathcal S^{1/2}(x)\,\phi_\ast(\xi),
\end{eqnarray}
where $\phi(x)=\phi[\mu^{-1}(\xi)]=[\phi\circ\mu^{-1}](\xi)\equiv\phi_\ast(\xi)$. It is obvious that $\phi_\ast$ is the representation of $\phi$ in the $\xi$-space.\\
\indent Now one can look for the appropriate factorization for $\hat h\equiv \hat h_0$, which admits the following factorization
\begin{eqnarray}\label{2.5}
\hat h=\hat q_+\hat q_-+\mathcal{E}_0,
\end{eqnarray}
where $\mathcal{E}_0$ is the ground-state energy and $\hat q_\pm$ be the following operators
\begin{eqnarray}\label{2.6}
\hat q_\pm=\mp\frac{\hbar}{\sqrt{2m_0}}\sqrt{U(x)}\frac{d}{dx}\sqrt{U(x)}+W(x).
\end{eqnarray}
where $W(x)$ is a superpotential. Then it is easy to verify that $\hat q_\pm$ and the Hamiltonian $\hat h$ satisfy the mutually commutation relations
\begin{alignat}{2}\label{2.7}
[\hat q_-,\hat q_+]&=\frac{2\hbar}{\sqrt{2m_0}}\,U(x)W'(x)\ \mathbb{\hat I}_\mathfrak{H}, \nonumber\\
[\hat h,\hat q_-]  &=-\frac{2\hbar}{\sqrt{2m_0}}\,U(x)W'(x)\,\hat q_-, \\
[\hat h,\hat q_+]  &=\frac{2\hbar}{\sqrt{2m_0}}\,\hat q_+U(x)W'(x), \nonumber
\end{alignat}
and this is nothing but the generally deformed oscillator algebra (GDOA) for PDEM, where the right-hand side of Eqs.~\eqref{2.7} is called the structure function of the algebra. \\
\indent In the $\xi$-representation, the PDEM operators $\hat q_\pm$ are expressed as
\begin{eqnarray}\label{2.8}
\hat q_\pm=\hat a_\pm\pm\frac{\hbar}{\sqrt{2m_0}}\frac{d}{d\xi}\ln\mathcal{S}^{1/2}_\ast(\xi),
\end{eqnarray}
where $\hat a_\pm$ are the conventional creation and annihilation operators of the quantum system with an arbitrary superpotential in the $\xi$-representation
\begin{eqnarray}\label{2.9}
\hat a_\pm=\mp\frac{\hbar}{\sqrt{2m_0}}\frac{d}{d\xi}+W_\ast(\xi).
\end{eqnarray}
\indent By acting the operators $\hat q_\pm$ on the left of $\mathcal{S}_\ast^{1/2}(\xi)$, we get
\begin{eqnarray}\label{2.10}
\hat q_\pm\mathcal {S}_\ast^{1/2}(\xi)=\mathcal {S}_\ast^{1/2}(\xi)\ \hat a_\pm,
\end{eqnarray}
and this simple result lead us to establish the action of $\hat q_\pm$ on the wave-function $\psi$ as
\begin{eqnarray}\label{2.11}
\hat q_\pm\psi(x)=\Big(\mathcal {S}_\ast^{1/2}(\xi)\ \hat a_\pm\Big)\phi_\ast(\xi).
\end{eqnarray}

\section{From GDOA to the $\mathfrak{su}(1,1)$ algebra}%

\noindent It is well-known that in the case of PDEM, the operators $\hat q_\pm$ do not act as ladder operators on the eigenfunctions unless if their commutator is reduced to a constant~\cite{34,39}. An interesting way to cope with this difficulty is to recover the $\mathfrak{su}(1,1)$ algebra from the GDOA by applying the similarity transformation~\cite{8} of the operators $\hat q_\pm$. An important implication of selecting new operators, say $\widehat Q_\pm$, other than $\hat q_\pm$ of GDOA for PDEM is that the irreducible representation of the $\mathfrak{su}(1,1)$ Lie algebra are formed by states which have the same energy eigenvalues as that of a \emph{certain} generalized shifted harmonic oscillator.\\
\indent In this way, to construct a well-defined $\mathfrak{su}(1,1)$ Lie algebra, we first introduce a similarity transformation\footnote{\;See Eqs.~(3.10) and (3.11) in Ref.~\cite{8}.} defined as
\begin{eqnarray}\label{3.1}
\hat q_\pm\ \rightarrow\ \widehat Q_\pm=(\hbar\omega_0)^{1/4}\,\mathcal{R}^{\mp 1/2}(x)\,\hat q_\pm\mathcal{R}^{\pm 1/2}(x),\qquad \textrm{and}\qquad \hat h\ \rightarrow\ \widehat H,
\end{eqnarray}
or equivalently
\begin{eqnarray*}
\mathcal{R}^{\pm 1/2}(x)\,\widehat Q_\pm=(\hbar\omega_0)^{1/4}\hat q_\pm\mathcal{R}^{\pm 1/2}(x),
\end{eqnarray*}
where $\widehat Q_\pm^\dagger=\widehat Q_\mp$. Here $\mathcal{R}(x)=1/r(x)$ and $r(x)$ is an unknown function to be determined.\\
\indent The next step is the construction of new creation and annihilation operators, which form a closed algebra. For this purpose, let us define
\begin{eqnarray}\label{3.2}
\widehat Q_\pm\ \rightarrow\ \mathcal{\widehat Q}_\pm=\frac{1}{\hbar\omega_0}\,\widehat Q_\pm\!^2,\qquad\textrm{and}\qquad \widehat H\ \rightarrow\ \mathcal{\widehat H},
\end{eqnarray}
with $\mathcal{\widehat Q}_\pm^\dagger=\mathcal{\widehat Q}_\mp$. This realization, Eqs.~\eqref{3.2}, is constructed in analogy with the so-called amplitude-squared realization of $\mathcal{SU}(1,1)$ group in quantum optics~\cite{31}, in which the ladder operators of the algebra are quadratic in the annihilation and creation operators. To clarify further, our global construction is characterized mathematically by identifying a set $\big\{\mathcal{\widehat H},\,\mathcal{\widehat Q_\pm}\big\}$ in the way that $\mathcal{\widehat H}$ is proportional to $\mathcal{\widehat K}_0$, while $\mathcal{\widehat Q}_+$ and $\mathcal{\widehat Q}_-$ are raising and lowering ladder operators, respectively; i.e.,
\begin{eqnarray}\label{3.3}
\mathcal{\widehat H}=c_0\mathcal{\widehat K}_0,\qquad \textrm{and}\qquad
\mathcal{\widehat Q_\pm}=c_\pm\mathcal{\widehat K_\pm},
\end{eqnarray}
so that the standard commutation relationships satisfy the Lie algebra corresponding to the $\mathcal{SU}(1,1)$ group
\begin{eqnarray}\label{3.4}
\big[\mathcal{\widehat K}_0,\,\mathcal{\widehat K_\pm}\big]=\pm\mathcal{\widehat K_\pm},\qquad\textrm{and}\qquad
\big[\mathcal{\widehat K}_-,\,\mathcal{\widehat K}_+\big]=2\mathcal{\widehat K}_0,
\end{eqnarray}
where $c_{0,\pm}$ are some constants to be determined.\\
\indent The associated Casimir operator is $\mathcal{\widehat C}^2\equiv\mathcal{\widehat K}_\pm\mathcal{\widehat K}_\mp-\mathcal{\widehat K}_0(\mathcal{\widehat K}_0\mp1)=k(k-1)\,\mathbb{\hat I}_{\mathfrak H}$ and $k$ is the Bargmann index labeling the irreducible representation. Thus the corresponding Hilbert space $\mathfrak{H}_k$ is spanned by the complete orthonormal basis $|n,k\rangle(\equiv|\psi_n(x)\rangle)$.\\
\indent Thus one has to merely construct the operator $\mathcal{\widehat H}$ such that it is equal to the commutator of $\mathcal{\widehat Q}_-$ and $\mathcal{\widehat Q}_+$. The method consists in choosing the function $r(x)$ in such a way that $\mathcal{\widehat H}$ contains the same kind of terms, due to the similarity transformation, as those already present in Eq.~\eqref{2.2}. We start by substituting Eqs.~\eqref{2.6} and~\eqref{3.1} into Eq.~\eqref{3.2}, which can be used to derived the second-order differential operator and after some straightforward but lengthy calculation, the expression of $r(x)$ can be formally obtained in terms of $W(x)$ and $\mu(x)$. It is given by
\begin{eqnarray}\label{3.5}
r(x)=\exp\bigg\{\frac{m_0\omega_0}{4}\,\mu^2(x)-\lambda\,\mu(x)-\frac{2\sqrt{2m_0}}{\hbar}\int^{\mu(x)}
W(y)\,d\mu(y)\bigg\},
\end{eqnarray}
where $\lambda$ is some constant of integration which will be kept as a parameter in the remainder of the article. Then the commutator of $\mathcal{\widehat Q}_-$ and $\mathcal{\widehat Q}_+$ have the following form
\begin{eqnarray}\label{3.6}
\big[\mathcal{\widehat Q}_-,\,\mathcal{\widehat Q}_+\big]&=&\mathcal{\widehat H}\nonumber\\
&=&-\frac{\hbar^2}{2m_0}
\bigg(\sqrt{U(x)}\frac{d}{dx}\sqrt{U(x)}\bigg)^2+\mathcal{V}_{\rm SHO.}(x),
\end{eqnarray}
where $\mathcal{V}_{\rm SHO.}(x)$ can be expressed as
\begin{eqnarray}\label{3.7}
\mathcal{V}_{\rm SHO.}(x)&=&\mathcal{W}^2(x)-\frac{\hbar}{\sqrt{2m_0}}\,U(x)\mathcal{W'}(x)+\mathcal{E}_0 \nonumber\\
&=&\frac{m_0\omega_0^2}{2}\bigg(\frac{\mu(x)}{4}-\frac{\lambda\hbar}{2m_0\omega_0}\bigg)^2,
\end{eqnarray}
which can be looked as a mass-deformed version of the generalized shifted harmonic oscillator. Thus from the expression obtained in Eq.~\eqref{3.7}, we shall deduce that
\begin{eqnarray}\label{3.8}
\mathcal{W}(x)=\frac{\omega_0}{4}\sqrt{\frac{m_0}{2}}\,\mu(x)-\frac{\lambda\hbar}{2\sqrt{2m_0}},
\qquad\textrm{and} \qquad \mathcal{E}_0=\frac{\hbar\omega_0}{8}.
\end{eqnarray}
\indent  Due to Eqs.~\eqref{3.1} and \eqref{3.5}, the operators which fulfill relations~\eqref{3.2} can be written as
\begin{eqnarray}\label{3.9}
\mathcal{\widehat Q}_\pm=\frac{1}{\sqrt{\hbar\omega_0}}\bigg(\mp\frac{\hbar}{\sqrt{2m_0}}\sqrt{U(x)}\frac{d}{dx}\sqrt{U(x)}
+\mathcal{W}(x)\bigg)^2,
\end{eqnarray}
and applying Eq.~\eqref{3.6} on Eqs.~\eqref{3.3} and~\eqref{3.4} we find that the coefficients $c_{0,\pm}$ are given by $c_0=\hbar\omega_0/2$ and $c_\pm=\sqrt{\hbar\omega_0}/2$. Thus, we finally obtain the complete structure of the commutators
\begin{eqnarray}\label{3.10}
\big[\mathcal{\widehat Q}_-,\,\mathcal{\widehat Q}_+\big]=\mathcal{\widehat H}\qquad\textrm{and}\qquad \big[\mathcal{\widehat H},\,\mathcal{\widehat Q}_\pm\big]=\pm\frac{\hbar\omega_0}{2}\,\mathcal{\widehat Q}_\pm,
\end{eqnarray}
which close the $\mathfrak{su}(1,1)$ algebra.

\section{New $\mathcal{SU}(1,1)$ PDEM CS}%

\noindent Once our algebra, Eqs.~\eqref{3.10}, has been expressed appropriately in terms of new annihilation and creation operators, we now turn our attention, as for the harmonic oscillator, to built up the corresponding PDEM CS for the potential \eqref{3.7}.\\
\indent Here we are going to construct them in the sense of Barut-Girardello CS. It can be seen from Eqs.~\eqref{3.10} that the operators $\mathcal{\widehat Q}_-$ and $\mathcal{\widehat Q}_+$ connect states with $\mathcal{\widehat H}$ and the energy scale is shifted by $\hbar\omega_0/2$ units under the $\mathfrak{su}(1,1)$ algebra; which means that they transform $|n,k\rangle$ into $|n+1,k\rangle$ and $|n-1,k\rangle$ with the additional condition that $\mathcal{\widehat Q}_-$ annihilates $|0,k\rangle$.\\
\indent Thus its discrete representation are given following (see, e.g.~\cite{31}):
\begin{eqnarray}\label{4.1}
\mathcal{\widehat Q}_+|n,k\rangle&=&\frac{1}{2}\,\sqrt{\hbar\omega_0(n+1)(n+2k)}\,|n+1,k\rangle,\nonumber\\
\mathcal{\widehat Q}_-|n,k\rangle&=&\frac{1}{2}\,\sqrt{\hbar\omega_0\,n(n+2k-1)}\,|n-1,k\rangle,\\
\mathcal{\widehat H}|n,k\rangle&=&\frac{\hbar\omega_0}{2}\,(n+k)\,|n,k\rangle\nonumber,\qquad\qquad\qquad
\quad(n=0,1,2,\cdots),
\end{eqnarray}
where from the last relation, the involved spectrum is given by $\mathcal{E}_n=\frac{\hbar\omega_0}{2}\,(n+k)$.\\
\indent Comparing $\mathcal{E}_n$ with the ground-state energy in Eq.~\eqref{3.8} we deduce that the lowest weight characterizing the irreducible representation is here $k=1/4$. Thus in our opinion this result is obvious in the sense that an analogy with the amplitude-squared realization is taken into account~\cite{31}. In order to construct the PDEM CS for the potential~\eqref{3.7}, we need a ground-state solution $|0,k\rangle$ which is an eigenstate of the operator $\mathcal{\widehat Q}_-$.

\subsection{Construction of PDEM wave-functions}%

\noindent In order to relate the solution of wave-functions to the standard problem of Hermite polynomials, it is easy first to verify that the following equality holds:
\begin{eqnarray}\label{4.2}
\pm\,\sqrt{U(x)}\frac{d}{dx}\sqrt{U(x)}+\mathcal{\overline W}(x)=\pm\,
e^{\mp\mathfrak{W}(x)}\bigg(\sqrt{U(x)}\frac{d}{dx}\sqrt{U(x)}\bigg)\,e^{\pm\mathfrak{W}(x)},
\end{eqnarray}
where
\begin{eqnarray}\label{4.3}
\mathfrak{W}(x)=\int^{\mu(x)}\mathcal{\overline W}(y)d\mu(y),\qquad \textrm{and}\qquad\mathcal{\overline W}(x)=\frac{\sqrt{2m_0}}{\hbar}\,\mathcal{W}(x).
\end{eqnarray}
\indent Thus the corresponding ground-state wave-function is the solution of $\mathcal{\widehat Q}_-|0,k\rangle=0$ and on taking Eqs.~\eqref{3.9} and~\eqref{4.2} into account, the equation $\mathcal{\widehat Q}_-|0,k\rangle=0$ can be expressed as a second-order differential equation
\begin{eqnarray}\label{4.4}
e^{-\mathfrak{W}(x)}\sqrt{U(x)}\,\frac{d}{dx}\bigg[U(x)\frac{d}{dx}\bigg(
\sqrt{U(x)}\,e^{\mathfrak{W}(x)}\psi_0(x)\bigg)\bigg]=0,
\end{eqnarray}
that has two independent solutions generated according to the brackets in Eq.~\eqref{4.4}, i.e.
\begin{eqnarray}\label{4.5}
\sqrt{U(x)}\,e^{\mathfrak{W}(x)}\psi_0(x)=\textrm{const.},\qquad\textrm{and}\qquad
U(x)\frac{d}{dx}\bigg(
\sqrt{U(x)}\,e^{\mathfrak{W}(x)}\psi_0(x)\bigg)=\textrm{const.}
\end{eqnarray}
\indent A brief examination yields to correspond the first solution to $k=1/4$ and the second to $k=3/4$; this latter will be avoided if we restrict ourself to the $k=1/4$ case taken at the beginning. Then the only solution of Eq.~\eqref{4.5} can be formally obtained in terms of $\mu(x)$ is
\begin{eqnarray}\label{4.6}
|0,\tfrac{1}{4}\rangle=\mathcal{N}_0\frac{1}{\sqrt{U(x)}}\exp\bigg[-\frac{m_0\omega_0}{8\hbar}\,\mu^2(x)
+\frac{\lambda}{2}\,\mu(x)\bigg],
\end{eqnarray}
where $\mathcal{N}_0$ is the zeroth-order normalization constant. Similarly, from Eqs.~\eqref{4.1} and using Eq.~\eqref{3.9}, states $|n,\tfrac{1}{4}\rangle\big(\!=|\psi_n(x)\rangle\big)$ are given by
\begin{eqnarray}\label{4.7}
|n,\tfrac{1}{4}\rangle=\mathcal{N}_n\frac{(2/\hbar\omega_0)^n}{\sqrt{n!\big(\tfrac{1}{2}\big)_n}}
\bigg(\!-\frac{\hbar}{\sqrt{2m_0}}\sqrt{U(x)}\frac{d}{dx}\sqrt{U(x)}
+\mathcal{W}(x)\bigg)^{2n}|0,\tfrac{1}{4}\rangle,
\end{eqnarray}
where $(a)_n$ is the Pochhammer's notation $(a)_n=a(a+1)\cdots(a+n-1)$ and $(a)_0=1$. By making use of Eq.~\eqref{4.2} it is easy to reformulate Eq.~\eqref{4.7} so that
\begin{eqnarray}\label{4.8}
|n,\tfrac{1}{4}\rangle&=&\mathcal{N}_n\frac{\big(\hbar/m_0\omega_0\big)^n}{\sqrt{n!\big(\tfrac{1}{2}\big)_n}}\,
e^{\mathfrak{W}(x)}\bigg(\!\sqrt{U(x)}\frac{d}{dx}\sqrt{U(x)}\bigg)^{2n}\,e^{-\mathfrak{W}(x)}
|0,\tfrac{1}{4}\rangle,\nonumber\\
&=&\mathcal{N}_n\frac{\big(\hbar/m_0\omega_0\big)^n}{\sqrt{n!\big(\tfrac{1}{2}\big)_n}}\,
e^{\frac{m_0\omega_0}{8\hbar}\mu^2-\frac{\lambda}{2}\mu}
\bigg(\!\sqrt{U(x)}\frac{d}{dx}\sqrt{U(x)}\bigg)^{2n}\,
e^{-\frac{m_0\omega_0}{8\hbar}\mu^2+\frac{\lambda}{2}\mu}|0,\tfrac{1}{4}\rangle.
\end{eqnarray}
\indent However if we start with the appropriate generating function~\cite{40}
\begin{eqnarray*}
H_n(p-q)=e^{-(2pq-q^2)}\bigg(\frac{d}{dq}\bigg)^n\,e^{2pq-q^2},
\end{eqnarray*}
where $H_n$ is the Hermite polynomials, it is possible using Eq.~\eqref{4.2} to generalize it in exact analogy with Eq.~\eqref{4.8} in such a way that
\begin{eqnarray}\label{4.9}
\mathbb{H}_{2n}\bigg[\sqrt{\frac{2m_0\omega_0}{\hbar}}\bigg(\!\frac{\lambda\hbar}{2m_0\omega_0}
-\frac{\mu(x)}{4}\bigg)\bigg]=e^{-\big(-\frac{m_0\omega_0}{8\hbar}\mu^2+\frac{\lambda}{2}\mu\big)}
\bigg(\!\sqrt{U}\frac{d}{dx}\sqrt{U}\bigg)^{2n}
e^{-\frac{m_0\omega_0}{8\hbar}\mu^2+\frac{\lambda}{2}\mu},
\end{eqnarray}
holds, where $q=\sqrt{\frac{2m_0\omega_0}{\hbar}}\frac{\mu(x)}{4}$ and $p=\lambda\sqrt{\frac{\hbar}{2m_0\omega_0}}$, and $\mathbb{H}_{2n}$ are basically the even Hermite polynomials where the signs of all coefficients are made positive. Therefore the state space $\mathfrak{H}_{1/4}$ is the even Fock space with the orthonormal basis consisting of even number eigenstates $|n,\tfrac{1}{4}\rangle=|2n\rangle$.\\
\indent Thus we see that the eigenfunctions, Eq.~\eqref{4.8}, are given by
\begin{equation}\label{4.10}
\begin{split}
|n,\tfrac{1}{4}\rangle=\mathcal{N}_n\frac{\big(\hbar/m_0\omega_0\big)^n}{\sqrt{n!\big(\tfrac{1}{2}\big)_n}}\, \mathbb{H}_{2n}\bigg(\sqrt{\frac{m_0\omega_0}{8\hbar}}\,\mu(x)
-\lambda\sqrt{\frac{\hbar}{2m_0\omega_0}}\,\bigg)\frac{1}{\sqrt{U(x)}} \,e^{-\frac{m_0\omega_0}{8\hbar}\mu^2(x)+\frac{\lambda}{2}\mu(x)},
\end{split}
\end{equation}
and to calculate the normalization constant $\mathcal{N}_n$, we are making use of Eq.~(\textbf{7.374} 2) of Ref.~\cite{41}. Thus the normalization constant is reduced to
\begin{eqnarray}\label{4.11}
\mathcal{N}_n=e^{-\frac{\lambda^2\hbar}{2m_0\omega_0}}
\sqrt{\bigg(\frac{m_0\omega_0}{2\hbar}\bigg)^{2n+1/2}\frac{n!\,\Gamma(n+1/2)}{\sqrt{2\pi}
\,\Gamma(2n+1/2)}},
\end{eqnarray}
and the eigenfunctions~\eqref{4.10} are expressed as
\begin{eqnarray}\label{4.12}
|n,\tfrac{1}{4}\rangle&=&\sqrt{\sqrt{\frac{m_0\omega_0}{\hbar}}\frac{1}{2^{2n+1}\Gamma(2n+\tfrac{1}{2})}}\, \mathbb{H}_{2n}\bigg(\sqrt{\frac{m_0\omega_0}{8\hbar}}\,\mu(x)
-\lambda\sqrt{\frac{\hbar}{2m_0\omega_0}}\,\bigg)\nonumber\\
&&\times\frac{1}{\sqrt{U(x)}}\,\exp{\bigg[-\bigg(\sqrt{\frac{m_0\omega_0}{8\hbar}}\,\mu(x)
-\lambda\sqrt{\frac{\hbar}{2m_0\omega_0}}\,\bigg)^2\Bigg]}.
\end{eqnarray}

\subsection{Construction of PDEM Barut-Girardello CS}%

\noindent The Barut-Girardello coherent states $|\alpha;k\rangle\big(\!\equiv|\Xi_\alpha(x)\rangle$\big) are defined as the eigenstates of the lowering operator $\mathcal{\widehat K_-}$. For the irreducible representation $k=1/4$, we have
\begin{eqnarray}\label{4.13}
\mathcal{\widehat K_-}|\alpha;\tfrac{1}{4}\rangle=\alpha|\alpha;\tfrac{1}{4}\rangle,
\end{eqnarray}
where $\alpha$ is an arbitrary complex number. Using Eqs.~\eqref{3.3} and~\eqref{4.1} we can obtain the Barut-Girardello CS as a linear combination of the orthonormal state basis $|n,\tfrac{1}{4}\rangle$
\begin{eqnarray}\label{4.14}
|\alpha;\tfrac{1}{4}\rangle=\sqrt{\frac{\sqrt{\pi}}{\cosh2|\alpha|}}\sum_{n=0}^\infty
\frac{\alpha^n}{\sqrt{n!\Gamma(n+\tfrac{1}{2})}}\,|n,\tfrac{1}{4}\rangle.
\end{eqnarray}
\indent Then inserting Eq.~\eqref{4.12} into Eq.~\eqref{4.14} we get the following PDEM Barut-Girardello CS in terms of the even Hermite polynomials
\begin{eqnarray}\label{4.15}
|\Xi_\alpha(x)\rangle&=&\sqrt{\sqrt{\frac{m_0\omega_0}{\hbar}}\frac{\sqrt{\pi/2}}{\cosh2|\alpha|}}
\frac{1}{\sqrt{U(x)}}\exp\bigg[-\bigg(\sqrt{\frac{m_0\omega_0}{8\hbar}}\,\mu(x)
-\lambda\sqrt{\frac{\hbar}{2m_0\omega_0}}\bigg)^2\bigg]\nonumber\\
&&\times\sum_{n=0}^\infty\frac{(\alpha/2)^n}{\sqrt{n!\Gamma(n+\tfrac{1}{2})
\Gamma(2n+\tfrac{1}{2})}}
\,\mathbb{H}_{2n}\bigg(\sqrt{\frac{m_0\omega_0}{8\hbar}}\,\mu(x)
-\lambda\sqrt{\frac{\hbar}{2m_0\omega_0}}\,\bigg).
\end{eqnarray}

\subsubsection{Mean values of the Hamiltonian $\mathcal{\widehat H}$}%

\noindent Now we can calculate easily the mean values $\langle\mathcal{\widehat H}\rangle_\alpha$ and $\langle\mathcal{\widehat H}^2\rangle_\alpha$, as well as its mean square deviation $(\Delta\mathcal{\widehat H})^2$ in a given PDEM CS $|\alpha;\tfrac{1}{4}\rangle$ of Eq.~\eqref{4.15}. To this end, let us evaluate first the distribution of finding the states $|\alpha;\tfrac{1}{4}\rangle\big(\!\equiv|\Xi_\alpha(x)\rangle\big)$ in the $|n,\tfrac{1}{4}\rangle\big(\!\equiv|\psi_n(x)\rangle\big)$ basis. It is given by the following identity
\begin{eqnarray}\label{4.16}
\mathfrak{D}_n[\Xi_\alpha]\ \stackrel{\mbox{\tiny def.}}{=}\ |\langle\psi_n(x)\,|\,\Xi_\alpha(x)\rangle|^2,
\end{eqnarray}
where making use of Eq.~\eqref{4.14} one obtain
\begin{eqnarray}\label{4.17}
\langle\psi_n(x)\,|\,\Xi_\alpha(x)\rangle&=&\sqrt{\frac{\sqrt{\pi}}{\cosh2|\alpha|}}
\sum_{m=0}^\infty\frac{\alpha^m}{\sqrt{m!\Gamma(m+\tfrac{1}{2})}}\,\langle\psi_n(x)\,|\,\psi_m(x)\rangle\nonumber\\
&=&\sqrt{\frac{\sqrt{\pi}}{\cosh2|\alpha|}}\frac{\alpha^n}{\sqrt{n!\Gamma(n+\tfrac{1}{2})}},
\end{eqnarray}
and then the distribution~\eqref{4.16} is given following
\begin{eqnarray}\label{4.18}
\mathfrak{D}_n[\Xi_\alpha]=\frac{\sqrt{\pi}}{n!\Gamma(n+\tfrac{1}{2})}
\frac{|\alpha|^{2n}}{\cosh2|\alpha|}.
\end{eqnarray}
\indent It is easy to verify that the probability of getting $\mathcal{E}_n=\frac{\hbar\omega_0}{2}(n+\frac{1}{4})$ as a result of a measurement of the energy is given by
\begin{eqnarray}\label{4.19}
\langle\mathcal{\widehat H}\rangle_\alpha&=&\langle\Xi_\alpha(x)\,|\,\mathcal{\widehat H}\,|\,\Xi_\alpha(x)\rangle\nonumber \\
&=&\sum_{n=0}^\infty\mathcal{E}_n\,\mathfrak{D}_n[\Xi_\alpha]\nonumber \\
&=&\frac{\hbar\omega_0}{2}\frac{\sqrt{\pi}}{\cosh2|\alpha|}\sum_{n=0}^\infty\frac{n+\tfrac{1}{4}}
{n!\Gamma(n+\tfrac{1}{2})}|\alpha|^{2n}\nonumber \\
&=&\frac{\hbar\omega_0}{8}\Big(1+4|\alpha|\tanh2|\alpha|\Big).
\end{eqnarray}
\indent It is worth noting that when $|\alpha|\ll1$, then $\langle\mathcal{\widehat H}\rangle_\alpha\simeq\mathcal{E}_0=\frac{\hbar\omega_0}{8}$. This shows that $(\Delta\mathcal{E})_\textrm{min}=\mathcal{E}_0$, the ground-state energy, belongs to the set of a such PDEM CS. We conclude that $|\Xi_\alpha(x)\rangle$ is the associated PDEM CS for the potential~\eqref{3.7} and is realized by the lowest energy state when the condition $|\alpha|\ll1$ is fulfilled.\\
\indent On the other hand the mean value of the quadratic operator, $\langle\mathcal{\widehat H}^2\rangle_\alpha$, can be obtained similarly, i.e.
\begin{eqnarray}\label{4.20}
\langle\mathcal{\widehat H}^2\rangle_\alpha&=&\langle\Xi_\alpha(x)\,|\,\mathcal{\widehat H}^2\,|\,\Xi_\alpha(x)\rangle\nonumber \\
&=&\sum_{n=0}^\infty\mathcal{E}_n^2\,\mathfrak{D}_n[\Xi_\alpha]\nonumber \\
&=&\frac{\hbar^2\omega_0^2}{64}\Big(1+16|\alpha|^2+16|\alpha|\tanh2|\alpha|\Big).
\end{eqnarray}
\indent Having calculated Eqs.~\eqref{4.19} and \eqref{4.20}, we can pass to evaluate the mean square deviation. Hence,
\begin{eqnarray}\label{4.21}
(\Delta\mathcal{\widehat H})^2&=&\langle\mathcal{\widehat H}^2\rangle_\alpha-\langle\mathcal{\widehat H}\rangle^2_\alpha\nonumber \\
&=&\frac{\hbar^2\omega_0^2}{64}\Big(8|\alpha|\tanh2|\alpha|+16|\alpha|^2(1-\tanh^22|\alpha|)\Big).
\end{eqnarray}
\indent For very large $|\alpha|$, we get $(\Delta\mathcal{\widehat H})_\infty\simeq\frac{\hbar\omega_0}{2}\sqrt{\frac{|\alpha|}{2}}$ and $\langle\mathcal{\widehat H}\rangle_\infty\simeq\frac{\hbar\omega_0}{2}|\alpha|$; hence $(\Delta\mathcal{\widehat H})_\infty\ll\langle\mathcal{\widehat H}\rangle_\infty\ $ which prove that, as usual for the Glauber states, the relative value of the energy of the state $|\Xi_\alpha(x)\rangle$ is well defined. Finally one may prove that the states $|\Xi_\alpha(x)\rangle$ minimize the generalized position-momentum uncertainty relation. As it is well-known, and from the $\mathcal{SU}(1,1)$ group ladder operators $\mathcal{\widehat K}_\pm$, we define
\begin{eqnarray}\label{4.22}
\mathfrak{\widehat X}=\frac{1}{2}\,\big(\mathcal{\widehat K}_++\mathcal{\widehat K}_-\big),\qquad\textrm{and}\qquad
\mathfrak{\widehat P}=\frac{i}{2}\,\big(\mathcal{\widehat K}_+-\mathcal{\widehat K}_-\big),
\end{eqnarray}
where, with these expressions, the operators $\mathfrak{\widehat X}$ and $\mathfrak{\widehat P}$ seem: (i) to satisfy the commutator $[\mathfrak{\widehat X},\,\mathfrak{\widehat P}]=i\mathcal{\widehat K}_0$, and (ii) minimizing the generalized position-momentum uncertainty relation
\begin{eqnarray}\label{4.23}
\Delta\mathfrak{\widehat X}\,\Delta\mathfrak{\widehat P}\geq\frac{1}{\hbar\omega_0}|\langle\mathcal{\widehat H}\rangle_\alpha|.
\end{eqnarray}
\subsubsection{Time evolution of PDEM CS}%

\noindent Due to the fact that the deduced energy eigenvalues $\mathcal{E}_n$ are equally spaced for $n=0,1,2,\cdots$, we argue that these states, $|\Xi_\alpha(x)\rangle$, also evolve in time without dispersion~\cite{42}. We are able to consider the dynamical evolution of PDEM CS in the $|n,\tfrac{1}{4}\rangle$ basis.\\
\indent The time evolution operator $\mathfrak{\widehat U}(t)$ for any Hamiltonian is defined as $\mathfrak{\widehat U}(t)=\exp[-\frac{i\mathcal{\widehat H}t}{\hbar}]$. In our case by acting on $|\Xi_\alpha(x)\rangle$, it turns out that
\begin{eqnarray}\label{4.24}
\mathfrak{\widehat U}(t)|\Xi_\alpha(x)\rangle&=&\sqrt{\frac{\sqrt{\pi}}{\cosh2|\alpha|}}\sum_{n=0}^\infty
\frac{\alpha^n}{\sqrt{n!\Gamma(n+\tfrac{1}{2})}}\,e^{-\frac{i\mathcal{\widehat H}t}{\hbar}}|\psi_n(x)\rangle,\nonumber \\
&=&\sqrt{\frac{\sqrt{\pi}}{\cosh2|\alpha|}}\sum_{n=0}^\infty
\frac{\alpha^n}{\sqrt{n!\Gamma(n+\tfrac{1}{2})}}\,e^{-i\frac{\omega_0t}{2}(n+\frac{1}{4})}
|\psi_n(x)\rangle,\nonumber \\
&=&e^{-i\frac{\omega_0t}{8}}|\Xi_\alpha(x,t)\rangle,
\end{eqnarray}
where $|\Xi_\alpha(x,t)\rangle\equiv|e^{-i\frac{\omega_0t}{2}}\,\Xi_\alpha(x)\rangle$. The temporal stability follows easily, which illustrates the fact that the time evolution of such coherent state remains within the family of coherent states.\\[-10mm]

\paragraph{\emph{Application}}%

\noindent Since the PDEM CS of the Eq.~\eqref{4.24} are labeled by the parameter $\alpha$ and depend explicitly on some appropriately deforming function $U(x)$, it is straightforward to choose these quantities in order to plot the squared modulus of $\Xi_\alpha(x,t)$.\\
\indent The simplest and particular choice is $U(x)=1$, where in this case $\mu(x)=x$. In the Figure 1 we plot probability densities $|\Xi_\alpha(x,t)|^2$ for the potential~\eqref{3.7} defined by $\lambda=2$ and given for the ten first states ($n_\textrm{max}=10$). We have taken three real values for $\alpha=0.5,\,1,\,2$, each of them at different time moments $t=3,\,5,\,7$ and we have used the atomic units $\hbar=\omega_0=2m_0=1$. It is clearly seen that $|\Xi_\alpha(x,t)|^2$ are localized in space and are all stable in different time moments, as for the harmonic oscillator.\\[3mm]

\begin{figure}[htbp]
\begin{center}
\includegraphics[width=15cm,height=4cm]{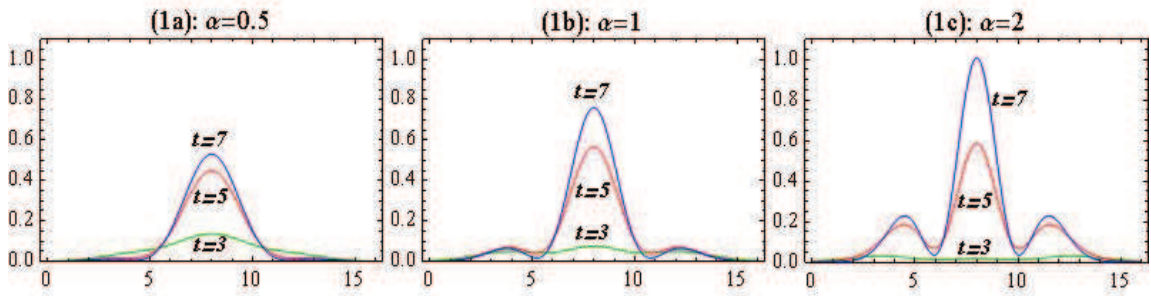}
\end{center}
\caption{\small{: Probability densities related to $\Xi_\alpha(x,t)$ for the parameters $\alpha=0.5,\,1,\,2$ at different time moments $t=3,\,5,\,7$.}}
\label{fig: cs2d}
\end{figure}

\indent In Figure 2 a three-dimensional plot of the squared modulus of $\Xi_\alpha(x,t)$ as a function of space-time is given for $\alpha=2,3,4$ and $\lambda=2$ showing that it leads to the well-behaved stability at all time. Moreover they oscillate back and forth as time goes by, which mimics the motion of the classical wave packet.\\[3mm]

\begin{figure}[h]
\begin{center}
\includegraphics[width=15.5cm,height=5cm]{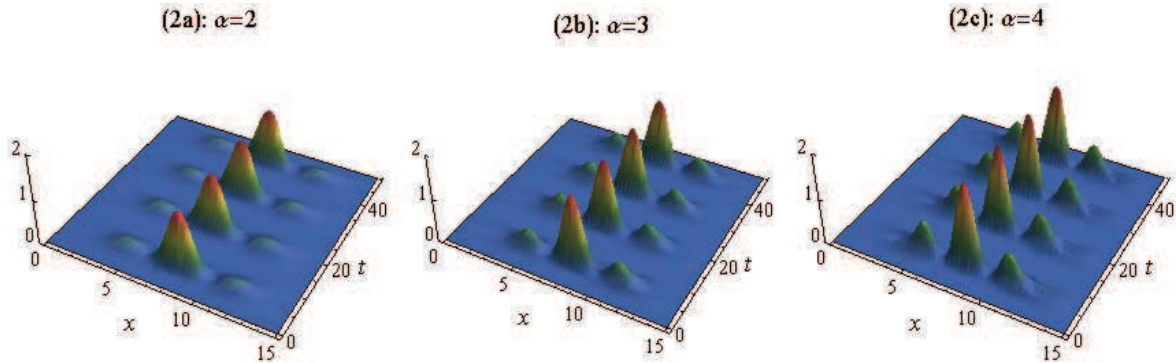}
\end{center}
\caption{\small{: Space-time evolution for the squared modulus of $\Xi_\alpha(x,t)$.}}
\label{fig: cs3d}
\end{figure}

\section{Conclusion}%

\noindent In this paper we have studied a new $\mathcal{SU}(1,1)$ PDEM CS related to the SHO potential. Starting from a similarity transformation, we have transformed the original set of generators $\big\{\hat q_\pm,\hat h\big\}$ associated to the GDOA for PDEM system to $\big\{\widehat{\mathcal Q}_\pm,\mathcal{\widehat H}\big\}$ related to the $\mathfrak{su}(1,1)$ Lie algebra. This allows us to tackle the problem of the generation of CS in a more straightforward and direct way and to obtain the well-known PDEM Barut-Girardello CS. We have deduced that the lowest weight characterizing the irreducible representation is $k=1/4$ mainly due, in our opinion, to the analogy with the amplitude-squared realization of $\mathcal{SU}(1,1)$ Lie group.\\
\indent Specifically, we have calculated the mean values of $\mathcal{\widehat H}$. The results demonstrate that the ground-state energy, $\mathcal{E}_0$, belongs to the set of PDEM CS when $|\alpha|\ll1$, and the relative value of the energy of the state $|\Xi_\alpha(x)\rangle$ is well defined for large $|\alpha|$ which agree, as usual, with the Glauber states. We have pointed out that the deduced PDEM CS minimize the generalized position-momentum uncertainty relation and preserve the same analytical form than those of Glauber states.\\
\indent We have also investigated the dynamical evolution of PDEM CS, which is quite simple due to the fact that the associated energy eigenvalues, $\mathcal{E}_n$, are equally spaced. In this context, we have displayed in figures the time evolution of probability densities which show that PDEM CS are localized and more stable with time evolution. Thus we have confirmed that CS probability densities oscillate back and forth as time goes by, which gives us the states with the similar properties than those of classical wave packets.


\end{document}